\documentstyle[twoside]{article}

\catcode`\@=11
\long\def\@makefntext#1{
\protect\noindent \hbox to 3.2pt {\hskip-.9pt
$^{{\eightrm\@thefnmark}}$\hfil}#1\hfill}               

\def\thefootnote{\fnsymbol{footnote}}
\def\@makefnmark{\hbox to 0pt{$^{\@thefnmark}$\hss}}    

\def\ps@myheadings{\let\@mkboth\@gobbletwo
\def\@oddhead{\hbox{}
\rightmark\hfil\eightrm\thepage}
\def\@oddfoot{}\def\@evenhead{\eightrm\thepage\hfil
\leftmark\hbox{}}\def\@evenfoot{}
\def\sectionmark##1{}\def\subsectionmark##1{}}

\oddsidemargin=\evensidemargin
\renewcommand{\thefootnote}{\fnsymbol{footnote}}

\newcounter{sectionc}\newcounter{subsectionc}\newcounter{subsubsectionc}
\renewcommand{\section}[1] {\vspace{12pt}\addtocounter{sectionc}{1}
\setcounter{subsectionc}{0}\setcounter{subsubsectionc}{0}\noindent
        {\tenbf\thesectionc. #1}\par\vspace{5pt}}
\renewcommand{\subsection}[1] {\vspace{12pt}\addtocounter{subsectionc}{1}
        \setcounter{subsubsectionc}{0}\noindent
        {\bf\thesectionc.\thesubsectionc. {\kern1pt \bfit #1}}\par\vspace{5pt}}
\renewcommand{\subsubsection}[1] {\vspace{12pt}\addtocounter{subsubsectionc}{1}
        \noindent{\tenrm\thesectionc.\thesubsectionc.\thesubsubsectionc.
        {\kern1pt \tenit #1}}\par\vspace{5pt}}
\newcommand{\nonumsection}[1] {\vspace{12pt}\noindent{\tenbf #1}
        \par\vspace{5pt}}

\newcounter{appendixc}
\newcounter{subappendixc}[appendixc]
\newcounter{subsubappendixc}[subappendixc]
\renewcommand{\thesubappendixc}{\Alph{appendixc}.\arabic{subappendixc}}
\renewcommand{\thesubsubappendixc}
        {\Alph{appendixc}.\arabic{subappendixc}.\arabic{subsubappendixc}}

\renewcommand{\appendix}[1] {\vspace{12pt}
        \refstepcounter{appendixc}
        \setcounter{figure}{0}
        \setcounter{table}{0}
        \setcounter{lemma}{0}
        \setcounter{theorem}{0}
        \setcounter{corollary}{0}
        \setcounter{definition}{0}
        \setcounter{equation}{0}
        \renewcommand{\thefigure}{\Alph{appendixc}.\arabic{figure}}
        \renewcommand{\thetable}{\Alph{appendixc}.\arabic{table}}
        \renewcommand{\theappendixc}{\Alph{appendixc}}
        \renewcommand{\thelemma}{\Alph{appendixc}.\arabic{lemma}}
        \renewcommand{\thetheorem}{\Alph{appendixc}.\arabic{theorem}}
        \renewcommand{\thedefinition}{\Alph{appendixc}.\arabic{definition}}
        \renewcommand{\thecorollary}{\Alph{appendixc}.\arabic{corollary}}
        \renewcommand{\theequation}{\Alph{appendixc}.\arabic{equation}}
        \noindent{\tenbf Appendix \theappendixc #1}\par\vspace{5pt}}
\newcommand{\subappendix}[1] {\vspace{12pt}
        \refstepcounter{subappendixc}
        \noindent{\bf Appendix \thesubappendixc. {\kern1pt \bfit #1}}
        \par\vspace{5pt}}
\newcommand{\subsubappendix}[1] {\vspace{12pt}
        \refstepcounter{subsubappendixc}
        \noindent{\rm Appendix \thesubsubappendixc. {\kern1pt \tenit #1}}
        \par\vspace{5pt}}

\newcommand{\textlineskip}{\baselineskip=13pt}
\newcommand{\smalllineskip}{\baselineskip=10pt}

\def\eightcirc{
\begin{picture}(0,0)
\put(4.4,1.8){\circle{6.5}}
\end{picture}}
\def\eightcopyright{\eightcirc\kern2.7pt\hbox{\eightrm c}}

\newcommand{\copyrightheading}[1]
        {\vspace*{-2.5cm}\smalllineskip{\flushleft
        {\footnotesize $\eightcopyright$\, World Scientific Publishing
         Company}\\
         }}

\def\abstracts#1#2#3{{
        \centering{\begin{minipage}{4.5in}\baselineskip=10pt\footnotesize
        \parindent=0pt #1\par
        \parindent=15pt #2\par
        \parindent=15pt #3
        \end{minipage}}\par}}

\newcommand{\bibit}{\nineit}
\newcommand{\bibbf}{\ninebf}
\renewenvironment{thebibliography}[1]
        {\frenchspacing
         \ninerm\baselineskip=11pt
         \begin{list}{\arabic{enumi}.}
        {\usecounter{enumi}\setlength{\parsep}{0pt}
         \setlength{\leftmargin 12.7pt}{\rightmargin 0pt} 
         \setlength{\itemsep}{0pt} \settowidth
        {\labelwidth}{#1.}\sloppy}}{\end{list}}

\newcounter{itemlistc}
\newcounter{romanlistc}
\newcounter{alphlistc}
\newcounter{arabiclistc}

\newcommand{\fcaption}[1]{
        \refstepcounter{figure}
        \setbox\@tempboxa = \hbox{\footnotesize Fig.~\thefigure. #1}
        \ifdim \wd\@tempboxa > 5in
           {\begin{center}
        \parbox{5in}{\footnotesize\smalllineskip Fig.~\thefigure. #1}
            \end{center}}
        \else
             {\begin{center}
             {\footnotesize Fig.~\thefigure. #1}
              \end{center}}
        \fi}

\newcommand{\tcaption}[1]{
        \refstepcounter{table}
        \setbox\@tempboxa = \hbox{\footnotesize Table~\thetable. #1}
        \ifdim \wd\@tempboxa > 5in
           {\begin{center}
        \parbox{5in}{\footnotesize\smalllineskip Table~\thetable. #1}
            \end{center}}
        \else
             {\begin{center}
             {\footnotesize Table~\thetable. #1}
              \end{center}}
        \fi}

\def\@citex[#1]#2{\if@filesw\immediate\write\@auxout
        {\string\citation{#2}}\fi
\def\@citea{}\@cite{\@for\@citeb:=#2\do
        {\@citea\def\@citea{,}\@ifundefined
        {b@\@citeb}{{\bf ?}\@warning
        {Citation `\@citeb' on page \thepage \space undefined}}
        {\csname b@\@citeb\endcsname}}}{#1}}

\newif\if@cghi
\def\cite{\@cghitrue\@ifnextchar [{\@tempswatrue
        \@citex}{\@tempswafalse\@citex[]}}
\def\citelow{\@cghifalse\@ifnextchar [{\@tempswatrue
        \@citex}{\@tempswafalse\@citex[]}}
\def\@cite#1#2{{$\null^{#1}$\if@tempswa\typeout
        {IJCGA warning: optional citation argument
        ignored: `#2'} \fi}}

\def\pmb#1{\setbox0=\hbox{#1}
        \kern-.025em\copy0\kern-\wd0
        \kern.05em\copy0\kern-\wd0
        \kern-.025em\raise.0433em\box0}

\def\fnt#1#2{\footnotetext{\kern-.3em
        {$^{\mbox{\scriptsize #1}}$}{#2}}}

\def\runninghead#1#2{\pagestyle{myheadings}
\markboth{{\protect\footnotesize\it{\quad #1}}\hfill}
{\hfill{\protect\footnotesize\it{#2\quad}}}}
\font\tenrm=cmr10
\font\tenit=cmti10
\font\tenbf=cmbx10
\font\bfit=cmbxti10 at 10pt
\font\ninerm=cmr9
\font\nineit=cmti9
\font\ninebf=cmbx9
\font\eightrm=cmr8

\def\qed{\hbox{${\vcenter{\vbox{                        
   \hrule height 0.4pt\hbox{\vrule width 0.4pt height 6pt
   \kern5pt\vrule width 0.4pt}\hrule height 0.4pt}}}$}}

\renewcommand{\thefootnote}{\fnsymbol{footnote}}        

\def\bsc{{\sc a\kern-6.4pt\sc a\kern-6.4pt\sc a}}       
\def\bflatex{\bf L\kern-.30em\raise.3ex\hbox{\bsc}\kern-.14em
T\kern-.1667em\lower.7ex\hbox{E}\kern-.125em X}

\begin{document}

\runninghead{Computation of coefficient functions}
{Computation of coefficient functions}

\normalsize\textlineskip
\thispagestyle{empty}
\setcounter{page}{1}

{\hfill Preprint INR-889/95}

\vspace*{0.88truein}

\centerline{\bf COMPUTATION OF COEFFICIENT FUNCTIONS FOR AN EXPANSION}
\vspace*{0.035truein}
\centerline{\bf OF PARAPOSITRONIUM DECAY WIDTH USING REDUCE
\footnote{
Talk presented at Fourth AIHENP95 Workshop, Pisa,
Italy, 3-8 April 1995}
}

\vspace*{0.37truein}
\centerline{\footnotesize A.A.PIVOVAROV
}
\vspace*{0.015truein}
\centerline{\footnotesize\it Institute for Nuclear Research}
\baselineskip=10pt
\centerline{\footnotesize\it Moscow, 117312,
Russia\footnote{E-mail: aapiv@ms2.inr.ac.ru}}
\vspace*{0.225truein}

\vspace*{0.21truein}

\abstracts{
A kind of asymptotic expansion for calculation
of the (para)positronium decay rate is formulated.
The method is completely relativistic and  gauge invariant.
The expansion goes over the inverse square of relative momentum $k$
of two photons in Euclidean region and the actual width is
determined at $k^2=-m^2$, $m$ being the electron mass.
}{}{}

\vspace*{1pt}\textlineskip      


\pagebreak

\textheight=7.8truein
\setcounter{footnote}{0}
\renewcommand{\thefootnote}{\alph{footnote}}

\noindent
At present, computation of physical observables with high precision
is of great importance for testing theory. In some cases the accuracy
of experimental data is better than theoretical predictions that
requires further calculations for adequate comparison.  The standard
perturbation theory can then become fairly difficult to live with.
Thus, the most recent data on lifetime of (ortho)positronium$^{1,2}$
seem to contradict the present theoretical calculations$^{3-5}$ done
in the next-to-leading order in fine structure constant $\alpha$.
Full two-loop calculations are not available yet and the discrepancy
is often parametrized by the value of second order coefficient (a
factor in front of $(\alpha/\pi)^2$) that should be about 250
(e.g.$^6$).  This number is considered quite big though not
inconceivable especially bearing in mind the large first coefficient.
Despite all speculations one cannot claim the real inconsistency
till the reliable computation of the second order coefficient and
confirmation of experimental measurements.  Recently another special
set of diagrams has been analytically computed in$^6$.  New
formulation based on explicit use of nonrelativistic effective theory
has been also implemented for getting a representation that stresses
the splitting of momentum region into relativistic and
nonrelativistic parts$^{7,8}$.  Along with these efforts
to have new additional
terms computed there are more formal attempts to resum the known
series using different kinds of summation technique, for instance,
Pade approximation.  Indeed, rewriting the series in equivalent form
up to relative order $\alpha$ one can make the required higher order
coefficients essentially smaller$^9$.
A remark is in order here.  Strictly speaking the rate should note be
represented as a power series in $\alpha$ -- even some analytical
functions of $\alpha$ connected with exact treatment of an initial
approximation can be kept in their concise form. This will not mean
any violation of consistency because $\alpha$ is not an expansion
parameter.$^{10}$ For fitting experiment, it implies that the change
of zero order approximation in Bethe-Salpeter approach
could bring new terms of any order
in $\alpha$ that can stand on the legal ground.  Still, it is
difficult to understand the required large value of the second order
coefficient.  The technique based on the Bethe-Salpeter equation
was widely used for computation of quarkonium decay rates$^{11,12}$
where a new nonrelativistic theory based approach has been recently
developed as well.$^{8,13}$

The present note describes a technique of calculating the
parapositronium decay rate into photons that is inspired by methods
of physics of strong interaction.$^{14}$ Results of explicit
calculation of parapositronium width up to the first order in
the fine structure constant are reported.$^{15}$

The Lehmann-Symanzik-Zimmermann reduction$^{16}$
leads to the following expression for the two-photon
decay rate of positronium
\begin{eqnarray}
\langle\gamma(k_1)\gamma(k_2)|P(p)\rangle=i(2\pi)^4\delta(p-k_1-k_2)e^2
\epsilon^{\mu}(k_1)\epsilon^{\nu}(k_2)\langle0|\Pi_{\mu\nu}(k)|P(p)\rangle,
\nonumber \\
T_{\mu\nu}(x)=iTJ_\mu(x/2)J_\nu(-x/2),
{}~~T_{\mu\nu}(k)=i\int TJ_\mu(x/2)J_\nu(-x/2)e^{ikx}dx
\label{1}
\end{eqnarray}
where $|P(p)\rangle$ is an approximation for the positronium bound
state, $p^2=m_P^2$, $k=(k_1-k_2)/2$, $J_\mu$ is the fermion
electromagnetic current, $\epsilon^{\mu}(k_1)$ is a polarization
vector of a photon.  Because of momentum conservation, $p=k_1+k_2$,
and mass-shell conditions for the photons, $k_1^2=k_2^2=0$, there
exist kinematical constraints on the variables $k$ and $p$, $kp=0$
and $k^2=-m_P^2/4$, $m_P$ being the "positronium" mass.

The main thing to analyze is therefore a $T$-product of two
electromagnetic currents $T_{\mu\nu}(k)$ that, being sandwiched
between the vacuum and positronium state, gives a matrix element of
the decay process
\begin{equation}
\langle 0|T_{\mu\nu}(k)|P(p)\rangle
=\epsilon_{\mu\nu\alpha\beta}k^\alpha p^\beta F(k^2),
{}~~~\Gamma(P\rightarrow\gamma\gamma)=|F(-{m_P^2\over 4})|^2
{e^4m_P^3\over 64\pi}.
\label{2}
\end{equation}

Little can be said about the vector $|P(p)\rangle$ besides that its
quantum numbers are $J^{PC}=0^{-+}$. It is quite analogous to the
neutral pion state in QCD. To the leading order in $\alpha$
Eq.~(\ref{2}) gives its decay width but corrections cannot be
unambiguously computed unless the electromagnetic contribution to the
pion wave function itself is properly taken into account. As for the
$T$-product of currents, it can be expanded in $x$ when $x$ goes to
zero.$^{17}$ For the matrix element, Eq.~(\ref{1}), this
expansion turns into a series in the ratio of two scales connected
with soft and hard parts of the process respectively. The first one
is the Bohr radius of the positronium ground state $r_B\sim 1/\alpha m$
that determines an effective size of the bound state wave function
and the second is the cutoff length set by the fermion mass $m$,
$r_{cut}\sim 1/m$. Eventually, it is an expansion in $r_{cut}/r_B$,
or in $\alpha$, but not completely since each term can contain some
higher powers of $\alpha$ as well.
The decay rate given by the form
factor $F(k^2)$ will be parametrized with some numbers that
describe the inner structure of positronium and can not be found
without analyzing its dynamics as a bound state.

To obtain the decay rate one needs the form factor
$F(k^2)$ only at the physical point $k^2=-m_P^2/4$.
The asymptotic expansion of the $T$-product can be done for any
value of $k^2$ in Euclidean region $k^2<0$ because
one does not encounter any physical singularities. Fortunately, the
physical point belongs to the allowed region and
one can compute the form factor
at the desired point $k^2=-m_P^2/4$. The convergence of
the expansion, which is asymptotic one, at this particular point is
not guaranteed.

Generating formula for the expansion reads
\begin{equation}
T_{\mu\nu}(x)
=\bar{\psi}(x/2)\gamma_\mu
G(x) \gamma_\nu \psi(-x/2)+ \bar{\psi}(-x/2)\gamma_\nu
G(-x) \gamma_\mu \psi(x/2)+\dots
\label{3}
\end{equation}
where $G(x)$ stands for a full fermion propagator.  Since one needs
only operators with parapositronium quantum numbers the $T$-product
can be antisymmetrized with respect to indices $\mu$ and $\nu$
(see Eq.~(\ref{2})). This is implicitly implied in the following.

The leading term has the form
\begin{equation}
T^0_{\mu\nu}(k)={2i\epsilon_{\mu\nu\alpha\beta}k^\alpha
\bar{\psi}\gamma^\beta \gamma_5 \psi \over m^2-k^2}
\label{4}
\end{equation}
and gives the contribution to the form factor $F(k^2)$
\begin{equation}
F^0(k^2)=-{2 \over m^2+\kappa^2}f_P,
\label{5}
\end{equation}
where $f_P$ is defined by the relation
$
\langle0|\bar{\psi}\gamma^\mu \gamma_5 \psi|P(p)\rangle=ip^\mu f_P,
$
and a notation $\kappa^2=-k^2=m_P^2/4$ is used. Clearly, this
quantity is similar to the pion decay constant.
After inserting a correction to the fermion propagator
$$
\Delta G(k)={e\tilde F_{\mu\nu}k^\mu\gamma^\nu \gamma_5\over
(m^2-k^2)^2}+{1\over 2}{em\sigma^{\mu\nu} F_{\mu\nu}\over
(m^2-k^2)^2},
{}~~\tilde F_{\mu\nu}={1\over
2}\epsilon_{\mu\nu\alpha\beta}F^{\alpha\beta},~\sigma^{\mu\nu}={i\over
2}[\gamma^\mu,\gamma^\nu] \nonumber
$$
into Eq.~(\ref{3}) and set arguments of fermion fields
equal to zero the following result for
a dimension-five operator connected with a photon emission follows
$$
T^F_{\mu\nu}(k)={2i\epsilon_{\mu\tau\nu\lambda}k_\rho
e\bar{\psi}\tilde F_{\rho\tau}\gamma^\lambda\psi \over
(m^2+\kappa^2)^2}.
$$
Defining the $\delta^2$ parameter by the relation
$$
\langle0|e\bar{\psi}\tilde F^{\rho\tau}\gamma^\lambda\psi|P(p)\rangle=
(p^\rho g^{\tau\lambda}-p^\tau g^{\rho\lambda})if_P{\delta^2/3}
$$
one finds
$$
F^F(k^2)={2 \over 3}{f_P\delta^2 \over(m^2+\kappa^2)^2}.
$$
The quantity $\delta^2$ is obviously of order $\alpha$ at least.

Next terms of the series are given by derivatives of fermion fields
after substitution of a free propagator $S(x)$ for the full one
$G(x)$. The first-order derivative gives a contribution
$$
T^1_{\mu\nu}(k)={2m k^\alpha
\bar{\psi}\sigma^{\mu\nu}
{{}^{{}^{{}^{\leftrightarrow}}}\!\!\!\!\!D}
_\alpha\psi \over (m^2+\kappa^2)^2}
$$
that turns into
$$
F^1(k^2)={f_P \over 3}{4m^2-m^2_P \over(m^2+\kappa^2)^2}
$$
after averaging
between the vacuum and positronium states. This term is also
suppressed by $\alpha$ due to presence of the difference
$4m^2-m^2_P$.

The second-order derivative leads to an operator
$\bar{\psi}\gamma^\lambda\gamma_5
{{}^{{}^{{}^{\leftrightarrow}}}\!\!\!\!\!D}
_\alpha
{{}^{{}^{{}^{\leftrightarrow}}}\!\!\!\!\!D}
_\beta\psi$,
the matrix element of which can be expressed through three numbers
A, B, C
$$
\langle0|\bar{\psi}
{{}^{{}^{{}^{\leftrightarrow}}}\!\!\!\!\!D}
^\alpha
{{}^{{}^{{}^{\leftrightarrow}}}\!\!\!\!\!D}
^\beta\gamma^\lambda\gamma_5\psi|P(p)\rangle
=if_P A g^{\alpha\beta}p^\lambda
+B(g^{\alpha\lambda}p^\beta+g^{\beta\lambda}p^\alpha)
+Cp^\alpha p^\beta p^\lambda$$
though only one of them contributes
$$
F^2(k^2)={f_PA \over(m^2+\kappa^2)^2}{3m^2+\kappa^2\over
m^2+\kappa^2}.
$$
I failed to show in general that the quantity $A$ is
of the order $\alpha$ in comparison with the leading
term, Eq.~(5).

The final representation up to dimension-five operators reads
$$
F(k^2)=-{2f_P \over m^2+\kappa^2}
(1-{\delta^2 \over 3(m^2+\kappa^2)}
-{4m^2-m^2_P\over 6(m^2+\kappa^2)}
-{A\over 2(m^2+\kappa^2)}{3m^2+\kappa^2\over m^2+\kappa^2}).
\eqno (6)
$$

For a consistent analysis perturbative corrections to the
coefficient function of the leading operator, Eq.~(\ref{4}), have to
be found. Being written in the form
$$
T^0_{\mu\nu}(k)={2i\epsilon_{\mu\nu\alpha\beta}k^\alpha
\bar{\psi}\gamma^\beta \gamma_5 \psi \over m^2-k^2}
(1+{\alpha\over 4\pi}c(z)),
\eqno (7)
$$
they are
$$
c(z)={c_1\over 3}(-z^2+10z+2+{12\over z+1}+{3\over
z})+{c_2\over 3}(z^2-7z-32+{12\over z})
$$
$$+{1\over
3}(-2z^2+19z-{11\over 2}+{24\over z+1}+{3\over z})
\eqno (8)
$$
where $z=\kappa^2/m^2$ and
$$
c_1=\int_0^1dx\ln(zx(1-x)+x)={(z+1)\ln(z+1)\over z}-2,
$$
$$
c_2={z\over 2}\int_0^1{dx\over zx(1-x)+1}={1\over a}\ln{a+1\over
a-1},~~~a=\sqrt{1+4/z}.
\eqno (9)
$$

The electron mass is the pole mass. The actual calculation
was performed in the $\overline{MS}$-scheme of renormalization with
its natural mass parameter $m_{\overline{MS}}$. A relation between
these two up to one-loop order is$^{18}$
$$
m=m_{\overline{MS}}\left(1+{\alpha\over \pi}(3\ln{\mu^2
\over m^2_{\overline{MS}}}+1)\right).
$$
The quantity $c(z)$ does not depend on the renormalization scheme if
the physical pole mass is used. The program of symbolic manipulations
REDUCE has been heavily used for computation.

To verify Eq.~(8) one can investigate two simple limits.
First, $z\rightarrow\infty$ that corresponds to $m\rightarrow 0$ or
$\kappa^2\rightarrow\infty$  at fixed  $m$.  In  this  limit  the
logarithmic  $z$-dependence  of  function  $c(z)$ must completely drop
out  because  of  the  zero  anomalous  dimension  of  the coefficient
function. The result is $c(\infty)=-7$. This result has been also
checked by independent calculation  with taking the  massless
limit from the very beginning. Though contributions of different
diagrams have been considerably rearranged a full agreement
for the final answer was found. The opposite limit $z\rightarrow 0$
also must not reveal any singularity.  It is easy to see that Eq.~(8)
is valid, $c(0)=7/2$.

Neglecting the difference between $4m^2$ and $m_P^2$, which is of
order $\alpha^2$, one gets
$$
1+{\alpha\over 4\pi}c(1)=1+1.012{\alpha\over 4\pi}.
$$

Now we briefly consider parameters of Eq.~(6). The amplitude $f_P$ is
connected with that of the positronium wave function at the
origin$^{19}$. The $\delta^2$ parameter is reduced to the matrix
element $\langle0|e\bar{\psi}\tilde
F_{\mu\nu}\gamma^\nu\psi|P(p)\rangle=ip_\mu f_P\delta^2$ that is
closely related to the corresponding $\delta^2$ parameter in  strong
interaction$^{20}$. As for the $A$ parameter, I did not  succeed in
expressing it in any transparent manner. Both $\delta^2$  and  $A$
parameters related to the positronium wave functions of nonleading
twists.

Eq.~(6) represents the main result. One sees that the
two-photon decay width of parapositronium is expressed through three
parameters $f_P$, $\delta^2$ and $A$
that require a special model to determine them.
Unambiguous physical meaning (up to a corresponding $\alpha$  order
within perturbative QED) can be given to physical
quantities that do not contain these parameters.
They
could be computed  independently if one knows the positronium wave
function from other sources (for instance, lattice calculation).
The main problem
how to define properly the vector $|P(p)\rangle$ cannot be answered
within the approach; so, strictly speaking, the model dependent
calculation of the parameters $f_P$, $\delta^2$ and $A$ are ambiguous
in higher orders in $\alpha$ though the parametrization (6) itself is
not. Another advantage of the proposed approach is a possibility to
sum up many corrections with the help of a nonlocal operator
expansion.$^{21-23}$

\nonumsection{Acknowledgements}
\noindent
Discussions with A.A.Penin are acknowledged.
I thank the organizers of AIHENP95 Workshop for
warm and creative atmosphere.
The work is supported in part by Russian Fund for Fundamental
Research under Grants Nos 93-02-14428 and 94-02-06427 and by
Soros Foundation.

\nonumsection{References}

\end{document}